\documentclass[a4paper,fleqn,usenatbib]{mnras}

\usepackage[T1]{fontenc}
\usepackage{ae,aecompl}

\usepackage{graphicx}	
\usepackage{amsmath}	
\usepackage{amssymb}	

\usepackage{color}
\usepackage[normalem]{ulem}

\title[Non-Transiting Exoplanets with Dusty Tails]{On the Detection of Non-Transiting Exoplanets with Dusty Tails}

\author[DeVore et al.]{
J. DeVore,$^{1}$\thanks{Email: devore@visidyne.com}
S. Rappaport,$^{2}$\thanks{Email: sar@mit.edu}
R.~Sanchis-Ojeda,$^{3,4}$\thanks{Email: sanchisojeda@berkeley.edu}
K. Hoffman,$^{5}$\thanks{Email:khoffman@seti.org}
and J. Rowe$^{6}$\thanks{Email:jason@astro.umontreal.ca}
\\
$^{1}$Visidyne, Inc., 111 South Bedford St., Suite 103, Burlington, MA 01803, USA; devore@visidyne.com\\
$^{2}$Department of Physics, and Kavli Institute for
  Astrophysics and Space Research, Massachusetts Institute of
  Technology, \\ Cambridge, MA 02139, USA; sar@mit.edu\\
$^{3}$Department of Astronomy, University of California, Berkeley, CA 94720, USA; sanchisojeda@berkeley.edu\\
$^{4}$NASA Sagan Fellow\\
$^{5}$SETI Institute, Mountain View, CA 94043, USA;khoffman@seti.org\\
$^{6}$Institut de recherche sur les exoplan\'etes, iREx, D\'epartement de physique, Universit\'{e} de Montr\'{e}al, QC, H3C 3J7, \\ Canada;jason@astro.umontreal.ca\\
}

\date{Accepted XXX. Received YYY; in original form ZZZ}

\pubyear{2016}

\begin{document}
\label{firstpage}
\pagerange{\pageref{firstpage}--\pageref{lastpage}}
\maketitle

\begin{abstract}
We present a way of searching for {\em non-transiting} exoplanets with dusty tails. In the transiting case, the extinction by dust during the transit removes more light from the beam than is scattered into it. Thus, the forward scattering component of the light is best seen either just prior to ingress, or just after egress, but with reduced amplitude over the larger peak that is obscured by the transit. This picture suggests that it should be equally productive to search for positive-going peaks in the flux from non-transiting exoplanets with dusty tails. We discuss what amplitudes are expected for different orbital inclination angles. The signature of such objects should be distinct from normal transits, starspots, and most - but not all - types of stellar pulsations.
\end{abstract}




\section{Introduction}

At the present time, there are three exoplanets which show fairly compelling evidence for emitting dusty effluents. These are KIC 12557548b (hereafter `KIC 1255b'; Rappaport et al.~2012); KIC 8639908 (KOI 2700b; Rappaport et al.~2014); and K2-22b (EPIC 201637175; Sanchis-Ojeda et al.~2015). The primary evidence for dusty tails being responsible for the transits includes: (1) highly and erratically variable transit depths (KIC 1255b and K2-22b); (2) a post transit depression in the flux (KIC 1255b and KOI 2700b); and (3) small positive bumps in the light curves just prior to ingress (KIC 1255b) or just after egress (K2-22b) that have been attributed to forward scattering when the bulk of the dust tail is off the disk of the host star.

These three exoplanet candidates have ultrashort orbital periods, $P_{\rm orb}$, i.e., shorter than 24 hours, e.g., 9.15 hr for K2-22b; 15.7 hr for KIC 1255b; and 22.1 hr for KOI 2700b. From a variety of arguments, they are taken to be small rocky planets with very low surface gravities, perhaps equal to that of Mercury or even lunar gravity (Rappaport et al.~2012; Perez-Becker \& Chiang 2013). The best explanation thus far to explain the loss of planetary mass in the form of dust is via a Parker-type wind of heavy metal vapors that condense into dust once the material has escaped the planet's gravity (Perez-Becker \& Chiang 2013).

If the interpretation of the pre-transit ``bump'' in the light curve of KIC 1255b (Rappaport et al.~2012; Brogi et al.~2012; Budaj 2013; van Werkhoven et al.~2014) and the post-transit ``bump'' in K2-22b (Sanchis-Ojeda 2015) as forward scattering from the dust tails is correct, then it follows that the highest amplitude portion of the forward scattering peak is obscured by the transit part of the light curve where extinction trumps forward scattering. This raises the intriguing possibility that {\em non-transiting} exoplanets with dusty tails might have a distinct and detectable signature via this same forward scattering peak. In this work we explore this possibility in some detail. We show that such a signature of dusty-tailed planets should be detectable with {\em Kepler}, K2, and TESS quality photometry (Borucki et al.~2010; Batalha et al.~2011; Howell et al.~2014; Ricker et al.~2014). 

The possibility of detecting non-transiting gas giant planets via reprocessing of stellar radiation from their atmospheres was discussed by Seager et al.~(2000) and Jenkins \& Doyle (2003). This work differs in that (i) very small (sub-Earth) non-transiting rocky planets may be detected in contrast to gas giants; (ii) it probes the dust emitted by the surface layers of the rocky planets, as opposed to the atmospheres of giant planets; and (iii) the forward scattering would be maximum at orbital phase $0^\circ$ as opposed to a maximum in the phase curve at $180^\circ$ in the case of gas giants. 

This work is organized as follows. In Sect.~\ref{sec:fs} we describe the basic model and its geometry. The effective scattering function, which is a convolution over the stellar disk of the stellar emittance with the phase function of the scattering dust particles, is computed in Sect.~\ref{sec:scat_prof} and Appendix A. The expected signals from forward scattering by dust tails in non-transiting planets is discussed in Sect.~\ref{sec:NTP}, and simulated light curves as a function of the orbital inclination angle are presented. Sect.~\ref{sec:detect} demonstrates how such signals could be readily detected with Fourier transforms. The detectability of such planet dust tails in non-transiting geometries with various past, present, and future missions is discussed in Sect.~\ref{sec:searches}. Finally, we conclude by suggesting why this idea may be important and what we can infer about the dust in the tails of non-transiting planets.

\section{The Forward Scattering Model}
\label{sec:fs}

The basic idea of the model is as follows. We assume there is a localized dust cloud near the planet which, during a transit, scatters some starlight out of the beam that was initially heading in the Earth's direction, but also scatters some starlight, not initially directed toward the Earth, into the observer's line of sight. Some of the basic geometry of this scattering is shown in Fig.~1. The top part of the figure simply illustrates how some of the photons initially heading toward the Earth are scattered out of the beam, whereas others are scattered into the line of sight. 

As we will show, for the case of an equatorial transit the attenuation of the stellar flux always exceeds the forward scattering component, though in some cases perhaps only by a factor of a few. 

We illustrate in Fig.~\ref{fig:K22b} an example of the relative contributions of the dust extinction vs.~the forward scattering enhancement for the case of the dusty-tailed planet K 22-b (Sanchis-Ojeda et al.~2015). The green curve is the model extinction curve as the dust cloud crosses the stelar disk, as compared to the blue curve which is the forward scattering contribution component of the model. The red curve, showing the net model transit profile, exhibits positive-going features only near the beginning and end of the transit\footnote{The positive displacement in the centroid of the forward scattering peak is due to the assumption that K2-22b has a {\em leading} dust tail rather than a trailing one (see Sanchis-Ojeda et al.~2015 for details).}. If there were no direct passage of the dust cloud over the stellar disk, only the forward scattering part, albeit with reduced amplitude, would remain. It is this forward scattering component which we examine in detail in the remainder of this work. 

\begin{figure}
\begin{center}
\includegraphics[width=0.48 \textwidth]{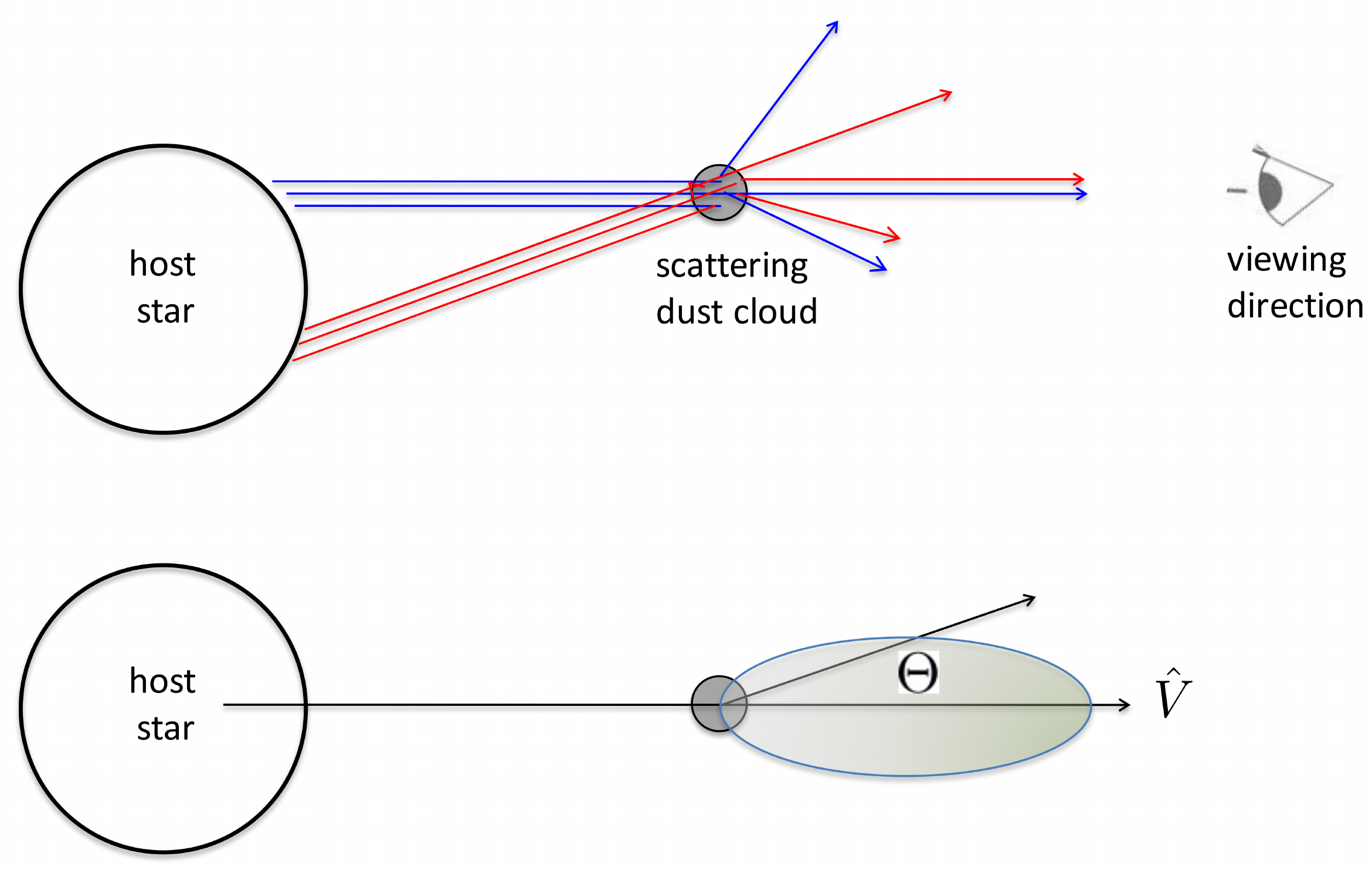}
\caption{Schematic dust-cloud scattering diagrams. The top sketch shows light from the host star (blue rays) initially headed in the direction of the observer but impinging on the scattering cloud; any rays that are scattered miss the observer. Other light rays (red) that are originally heading in directions other than the observer can impinge on the dust cloud, and some of these will scatter into the line of sight. The former effect always removes more light from the beam at the Earth, than is scattered back into the beam. The ratio of the two effects at midtransit is given by Eqn.~(\ref{eqn:scatter}). The bottom sketch defines the net effective beaming angle, $\Theta$, for the radiation leaving the cloud. It is a convolution of the angular distribution of starlight from the extended stellar surface and the scattering phase function, $P(\theta)$, of the dust particles (see Appendix \ref{app:ptheta}). }
\label{fig:scatter}
\end{center}
\end{figure}

\subsection{The Scattering Profile}
\label{sec:scat_prof}

The bottom panel in Fig.~1 defines the basic geometry of the model that we will utilize. The basic assumption used in this work is that the dust scattering cloud is spherical and small compared to the size of the host star. If the dust cloud is substantially elongated, then the results presented here need to be convolved with the density profile of the dust tail. However, at least for the three cases of dust-emitting planets we know of, the tail length is not significantly longer than the size of the host star\footnote{ See Table 6 in Sanchis-Ojeda et al.~2015 where the characteristic lengths of the dust tails in KIC 1255b and K2-22b are listed as $\sim$0.9 and $\lesssim 0.5$ times the host-star radius, respectively.}. In any case, our results are not substantially affected by the simplification of a small scattering cloud size, while the essence of the proposed idea should be considerably clarified.

Under the above assumptions, an effective scattering pattern, $\mathcal{P}(\Theta)$, can be defined as a convolution of the emission from the host star with the combined scattering phase function of all the different particles in the dust cloud, $P(\theta)$. 
In general, we expect the dust cloud to be comprised of a range of particle sizes, possibly described by a power-law or log-normal size distribution. For simplicity in this work, we consider only particles of a single size, but the results can be readily extended to any specific particle size distribution. We also take the particles to be spherical with radius, $a$. The angle $\Theta$ is defined as the angle between the line connecting the star and dust cloud, and the direction of the outgoing scattered radiation (see Fig. \ref{fig:scatter}). It is important to note that this geometry is independent of the observer's viewing angle. 

\begin{figure}
\begin{center}
\includegraphics[width=0.48 \textwidth]{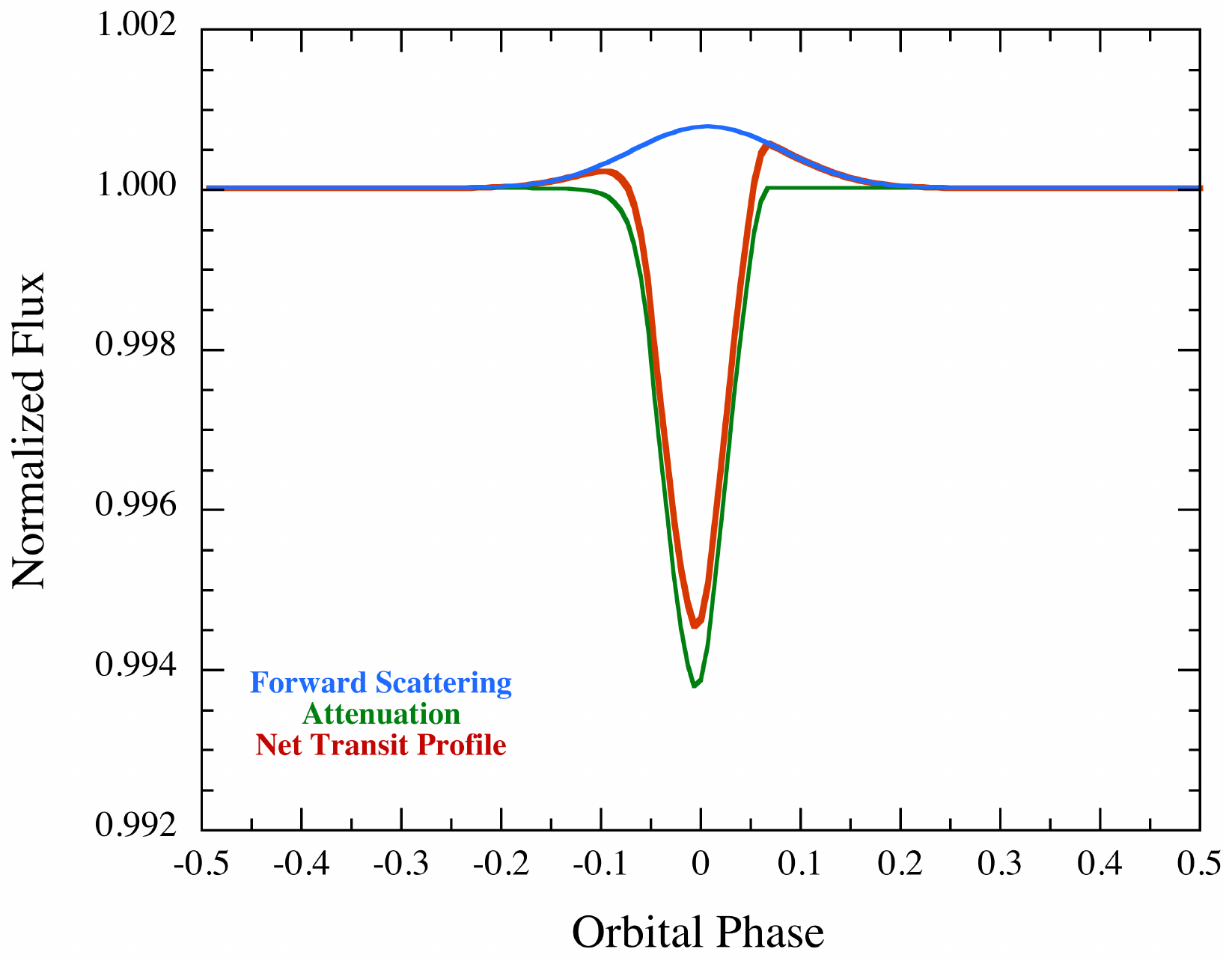}
\caption{Schematic of extinction of starlight vs.~the forward scattering of light into the beam. The blue forward scattering peak is what an observer would see in the absence of a transit, as long as the orbital inclination angle is not too small. Adapted from the model fit to the transit in K2-22b.}
\label{fig:K22b}
\end{center}
\end{figure}

We use the definition of the effective phase function given in Appendix A (Eqns.~\ref{eqn:ratio} and \ref{eqn:EPF}) to compute $\mathcal{P}(\Theta)$. Ideally, we should model the dust scattering with Mie scattering formalism. However, we have found that the Mie phase functions are modestly sensitive to the wavelength-dependent real and imaginary parts of the index of refraction. Therefore, to first introduce what the effective phase functions look like, and their progression with particle size, we illustrate the integrals of Eqns. ~\ref{eqn:ratio} and \ref{eqn:EPF} using simple Airy functions in lieu of Mie scattering. We return in Appendix A to show what these effective phase functions look like using Mie scattering. Previous work by two of the authors (DeVore et al.~2013) showed the benefits of modeling forward scattering using Airy diffraction, which is independent of the index of refraction. The simple Airy pattern we use is given by:
\begin{equation}
P(\theta) =  \frac{1}{\pi}\left[\frac{J_1(2 \pi a \sin \theta/\lambda)}{\sin \theta}\right]^2 ~~{\rm sr}^{-1}
\end{equation}
This is normalized to unity (to within 1.5\%) when integrated over solid angles in the forward hemisphere, as long as the ratio $a/\lambda \gtrsim 0.3$ where $\lambda$ is the observing wavelength. We find that this replicates the most important properties of Mie scattering. One caveat is that the Airy profile is not defined for scattering angles $\gtrsim 90^\circ$, so this eliminates any discussion of backscattered radiation. However, our Mie scattering calculations (see Appendix A) have informed us that this is not an important consideration for this problem (as long as the particles are relatively large, $a \gtrsim 0.3 \lambda$).

\begin{figure}
\begin{center}
\includegraphics[width=0.476 \textwidth]{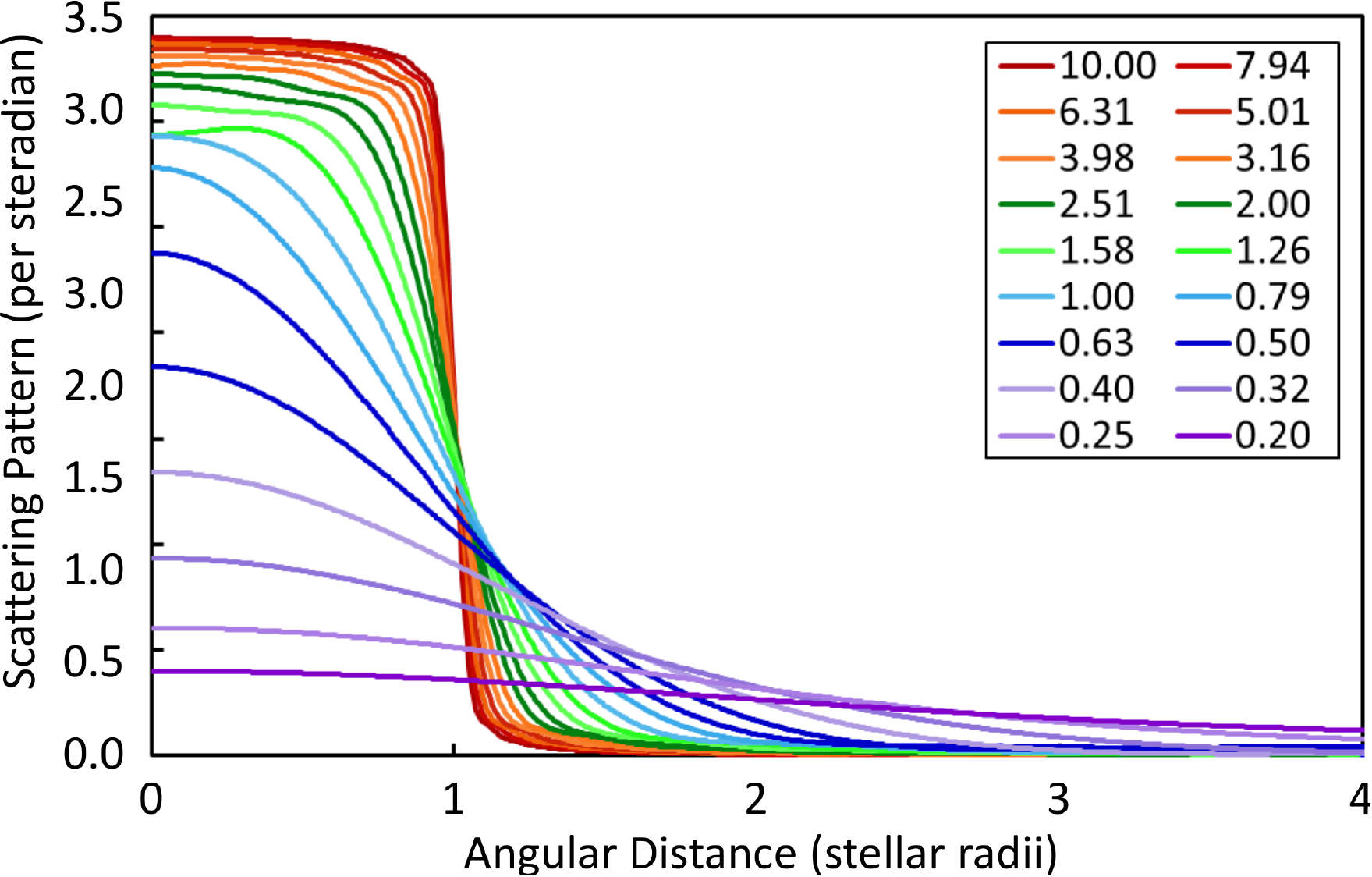} 
\includegraphics[width=0.476 \textwidth]{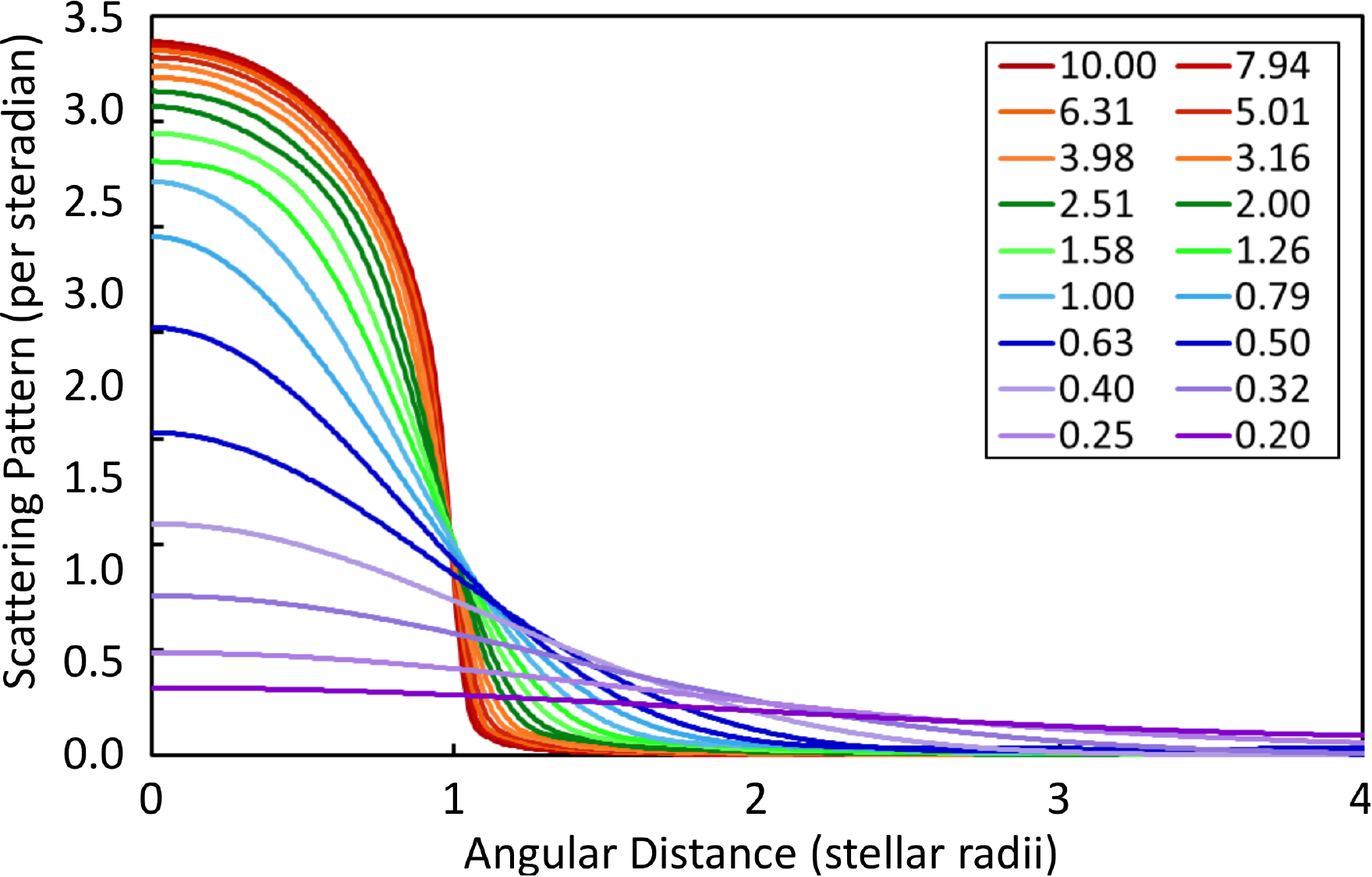} 
\caption{The effective scattering phase function, $\mathcal{P}(\Theta)$, i.e., the convolution of the angular distribution of starlight from the extended stellar surface and the scattering phase function $P(\theta)$, for a range of different grain sizes (see Appendix A for details). The top panel is for a Lambertian stellar surface brightness, while the bottom panel includes limb darkening for a typical K star. The color coding represents the grain size in microns, as given in the legend. The assumed observing wavelength is 0.55 $\mu$m. The reference angle to which the horizontal axis is scaled is the angular size of the star as viewed from the vantage point of the planet.}
\label{fig:phasefunc}
\end{center}
\end{figure}

An illustrative set of the net effective scattering profiles, $\mathcal{P}(\Theta)$, is shown in Fig.~\ref{fig:phasefunc} for different particle sizes. We took as a representative angular extent of the star, $\theta_s = 18^\circ$, which corresponds to a scaled stellar radius of $d/R_s = 3.3$ where $d$ is the orbital radius of the planet and $R_s$ is the radius of the host star. Note that $\theta_s$ is both the half angle of the transit time of a point source, and the angular extent of the star as seen from the vantage point of the planet. 

An inspection of Fig.~\ref{fig:phasefunc} shows how $\mathcal{P}(\Theta)$ goes from tracing the stellar disk radiation profile for large particles that have a narrow phase function, $P(\theta)$, to very broad $\mathcal{P}(\Theta)$ for small particles whose phase function is substantially broader than $\theta_s$. For a derivation of how $\mathcal{P}(\Theta)$ is calculated, as well as what the effective phase functions look like when $\mathcal{P}(\Theta)$ is computed with Mie scattering, see Appendix A. 

In order to get a feel for the characteristic angular size of the forward scattering peak of an individual dust grain, we note that for large grains, the shape of the scattering peak obtained from detailed Mie calculations is not very different from the Airy pattern of a circular aperture whose radius is $a$ (see DeVore et al.~2013; Appendix A). Thus, the characteristic (half) angle of the forward scattering is, to a good approximation, given by
\begin{equation}
\theta_{\rm scat} \approx \frac{0.61\lambda}{a} \simeq \frac{20^\circ}{a_{\mu{\rm m}}}
\label{eqn:angle}
\end{equation}
for visible radiation. The angle given by Eqn.~(\ref{eqn:angle}) typically encircles 90\% of the power in the Airy pattern. If we {\em require} 90\% encircled energy for all $\lambda$ and $a$ relevant to the problem, then $\theta_{\rm scatt}$ could be off by as much as 25\% by using Eqn.~(\ref{eqn:angle}). 

\subsection{Competition Between Forward Scattering and Extinction}
\label{sec:compet}

At the midpoint of an equatorial transit, the ratio of forward scattering to extinction by the dust is given by 
\begin{equation}
\frac{{\rm Scattering~Amplitude}}{{\rm Transit~Depth}} \simeq \pi \frac{\varpi \tau e^{-\tau}}{(1- e^{-\tau})} \,\left(\frac{R_s}{d}\right)^2\, \mathcal{P}(0)
\label{eqn:scatter}
\end{equation}
(see Appendix A), where $\tau$ is the optical depth of the scattering cloud (taken to be a sphere of uniform opacity everywhere), and $\varpi$ is the single scattering albedo (the ratio of the scattering to the extinction cross section). We have chosen to normalize both $P(\theta)$ and $\mathcal{P}(\Theta)$ over all solid angles to unity (rather than the conventional normalization of the scattering phase function, $P(\theta)$ to $4\pi$).

\subsection{Signal From a Non-Transiting Planet}
\label{sec:NTP}

\begin{figure}
\begin{center}
\includegraphics[width=0.48 \textwidth]{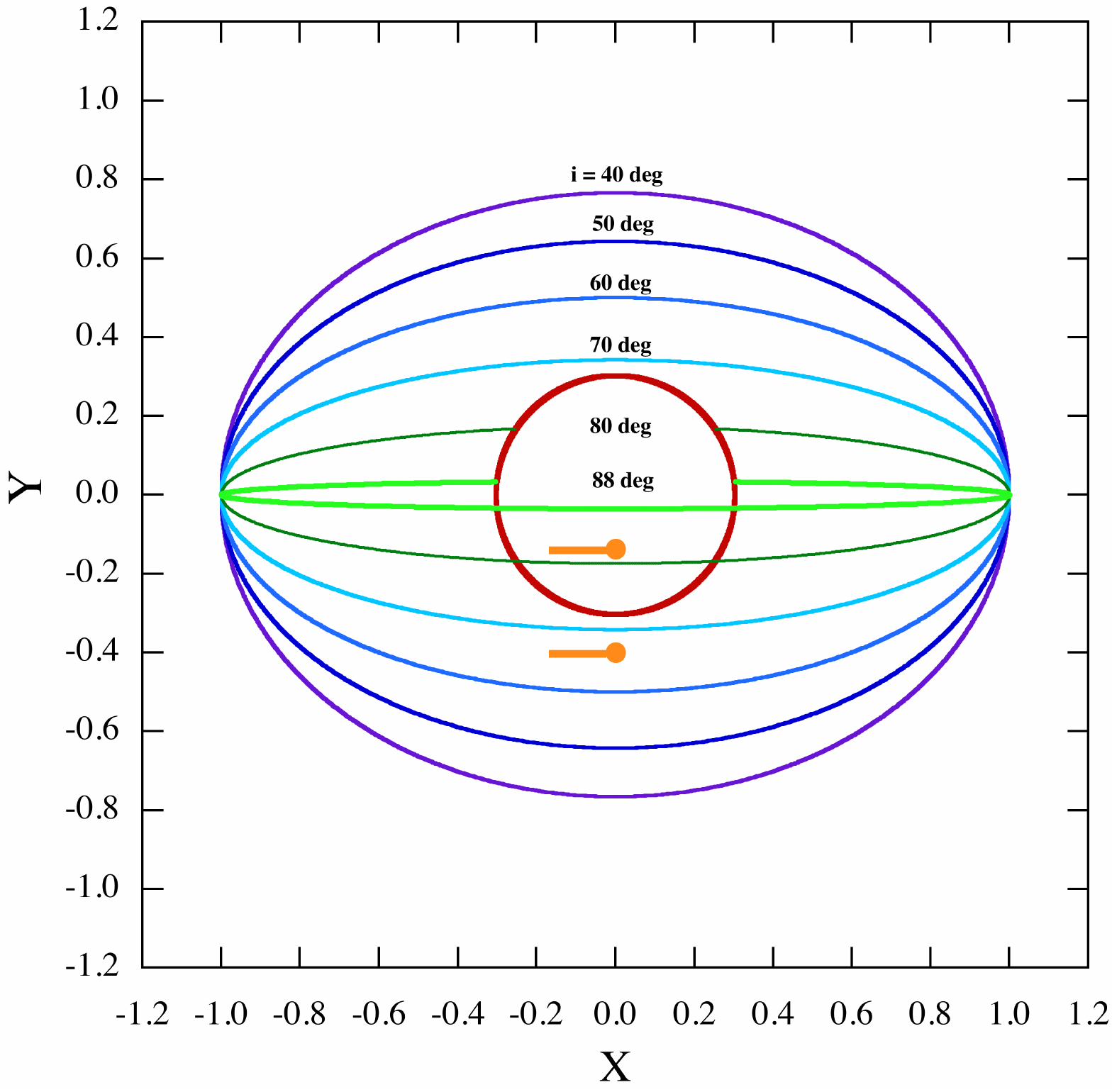}
\caption{Illustrative set of circular orbits as seen by a distant observer for 6 different orbital inclination angles. The host star is taken to have $d/R_s$ in this schematic of 3.3. Planets orbiting with $i > 72^\circ$ will exhibit transits of the host star. In those cases there will be transits with possible forward scattering peaks either before or after the transit (as are seen in KIC 1255b and K2-22b). For cases with smaller inclination angles, $i < 72^\circ$ the planet will not transit the host star, and only a forward-scattering peak will be present.}
\label{fig:orbits}
\end{center}
\end{figure}

Figure \ref{fig:orbits} shows the projections of several circular orbits for a range of orbital inclination angles from $i = 40^\circ$ to $88^\circ$. The central star has a value of $d/R_s = 3.3$. For $i \lesssim 73^\circ$ there will be no transits of the planet across the disk of the host star, and any signal from such a planet would likely come from forward scattering that peaks near orbital phase zero.

If we know the angular distribution, $\mathcal{P}(\Theta)$, from the combined finite angular size of the star, as viewed at the planet, with the dust scattering phase function, then we can compute the brightness of the scattered radiation at any point in an arbitrarily inclined orbit. We take the orbital phase to be $\phi \equiv \omega t$, where $\phi =0$ is defined as the phase where the planet is closest to the observer (mid-transit in the case where there are transits). With this simple geometric setup, the angle $\Theta$ where the observer samples the scattered radiation is given by:
\begin{equation}
\Theta = \cos^{-1}[\cos \phi \sin i]
\label{eqn:theta}
\end{equation}

If we also know the effective scattering function of the star and dust tail, then we can directly compute the intensity, $\mathcal{F}$, of the dust scattering as a function of the orbital phase:
\begin{equation}
\mathcal{F}(\phi) \simeq \pi \frac{\varpi \tau e^{-\tau}}{(1- e^{-\tau})} \,\left(\frac{R_s}{d}\right)^2\, \mathcal{P}(\cos^{-1}[\cos \phi \sin i])
\label{eqn:lightcurve}
\end{equation}
(see Appendix A). We use this expression to generate some illustrative light curves for forward scattering, by non-transiting dust emitting planets.

Examples of sequences of forward scattering peaks based on Eqn.~(\ref{eqn:lightcurve}) are shown in Fig.~\ref{fig:fs_peaks}. The simulated forward scattering peaks are shown for five orbital cycles, and, given the likely short orbital periods for these systems, this plot might represent a few days of the lightcurve. The simulated lightcurves are shown for four different orbital inclination angles, and one can see how dramatically the amplitudes of the forward scattering peaks diminish for decreasing inclination angles. This is due to the fact that most of the forward scattering pattern misses the Earth for low orbital inclination angles. The dust particle size adopted for this particular simulation was $a=0.3\,\mu$m. 
	
If we look at the absolute magnitudes of the simulated forward scattering peaks in Fig.~\ref{fig:fs_peaks}, they range from 300 ppm at $i = 70^\circ$ to 65 ppm at $i = 40^\circ$. If we adopt a lower limit of $\sim$100 ppm for the robust detectability of most {\em Kepler} targets with short orbital periods, then the range of inclination angles where the forward scattering peaks can be detected is in this relatively narrow range of $45^\circ \lesssim i \lesssim 70^\circ$. The interesting bottom line is that there is nearly the same solid angle for transiting systems as there is for potentially detectable non-transiting planets, e.g., with $i$ between $45^\circ$ and $70^\circ$ (see Sect.~\ref{sec:numbers}).

\subsubsection{Numbers of Transiting vs.~Non-Transiting Systems}
\label{sec:numbers}

Here we estimate the ratio of non-transiting systems that might be discovered via their forward scattering peaks to transiting systems with dusty tails which would be primarily detected via the extinction of starlight. We compute the ratio, $\mathcal{R}$, of weighted solid angles between close-in non-transiting systems with dust tails in orbits with $0^\circ <  i  \lesssim i_{\rm crit}$ and close-in transiting systems with $i_{\rm crit} \lesssim  i  \lesssim 90^\circ$, where $i_{\rm crit}$ is the critical orbital inclination angle for the occurrence of transits. This ratio can be expressed as:
\begin{equation}
\mathcal{R} \simeq \frac{\int_{0^\circ}^ {i_{\rm crit}} w(i) \, \sin i \,di} {\int_{i_{\rm crit}} ^{90^\circ} \sin i \,di} 
\end{equation}
where $w(i)$ is the weighting factor as a function of $i$ that takes into account the efficacy of forward scattering by the dust particles to significantly overlap our line of sight. The critical inclination angle for a transit is $i_{\rm crit} = \cos^{-1}(R_s/d)$. If we take some representative weighting factor, $\langle w \rangle$, to be a constant over the angular range 45$^\circ$ to $i_{\rm crit}$ and zero below 45$^\circ$, then the integral is analytic and leads to a simple result:
\begin{equation}
\mathcal{R} \simeq \langle w \rangle\left[ \frac{d}{\sqrt{2}R_s}-1 \right]   ~~.
\end{equation}
For illustrative values of $\langle w \rangle = 1$, and $d/R_s$ ranging from 3 to 6, the values of $\mathcal{R}$ range from 1.2 to 3.2.

The implication of all this is that there may be a comparable number of dusty-tailed planets that can potentially be detected via forward scattering only as there are via their transits. Of course, the exact ratio depends on the details of the angular structure of the forward scattered radiation and the search algorithms that are used to find such systems. We discuss these effects in the subsequent sections.

\section{Detection of Non-Transiting Dusty Planets}
\label{sec:detect}

\subsection{Fourier transforms}

\begin{figure}
\begin{center}
\includegraphics[width=0.48 \textwidth]{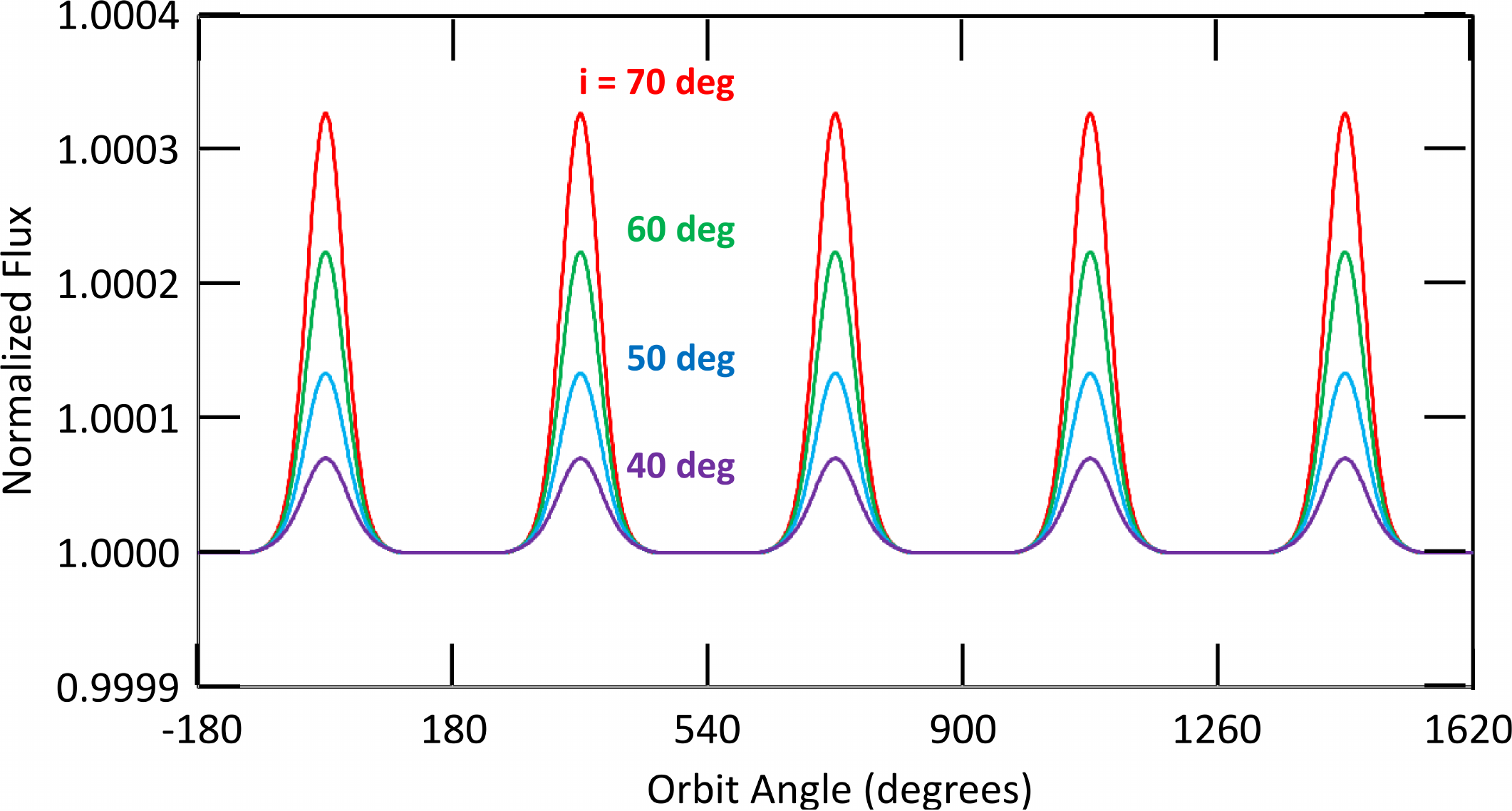}
\caption{Illustrative sequence of forward-scattering peaks for a non-transiting planet. The color coding of the curves indicates different orbital inclination angles ranging from barely missing the host star's disk (tallest red peaks) to orbits that are more tilted with respect to the observer's line of sight (small blue peaks). The assumed transit depth for all of these in transiting geometry would be 3000 ppm. The tallest peaks in non-transiting geometry (for $i = 70^\circ$) are 320 ppm.}
\label{fig:fs_peaks}
\end{center}
\end{figure}

\begin{figure}
\begin{center}
\includegraphics[width=0.48 \textwidth]{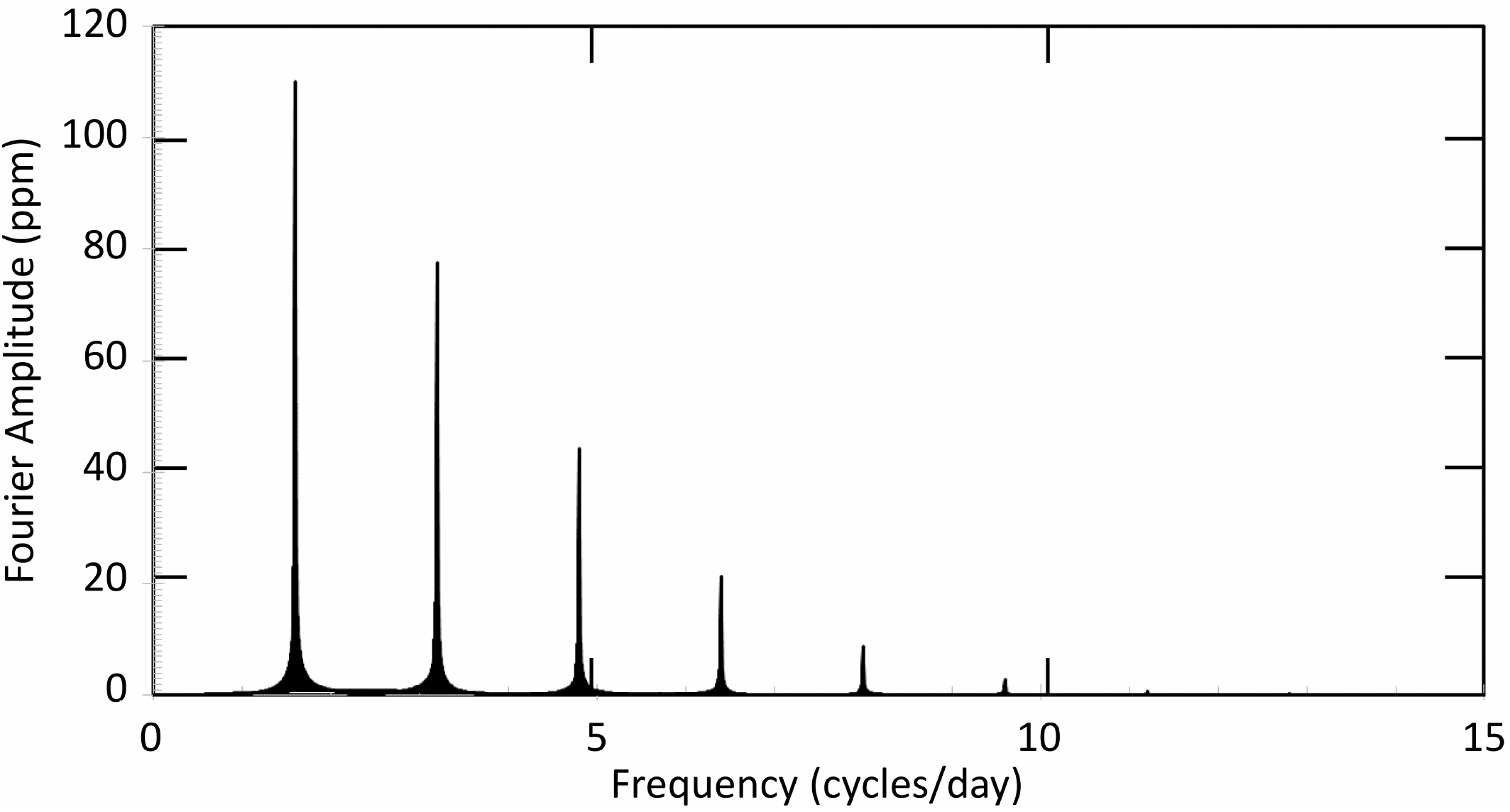}
\caption{Typical Fourier transform of a time series of forward scattering peaks from a non-transiting planet. The falloff in amplitude with increasing frequency is more rapid than would be the case for the corresponding dips of a transiting planet. However, they fall off more slowly than the FT amplitudes from starspots generally do. And, unlike in the case of starspots, the forward scattering peaks in flux have a long-term precise underlying frequency without phase drifts.}
\label{fig:FT}
\end{center}
\end{figure}

In Figure~\ref{fig:FT} we show the Fourier transform for the idealized forward scattering signal depicted in Fig.~\ref{fig:fs_peaks} for $i = 70^\circ$. If the expected forward-scattering peaks are sufficiently narrow (i.e., with respect to an orbital cycle), then the FT should exhibit a number of detectable harmonics -- as are seen in Fig.~\ref{fig:FT}. 

\begin{figure}
\begin{center}
\includegraphics[width=0.48 \textwidth]{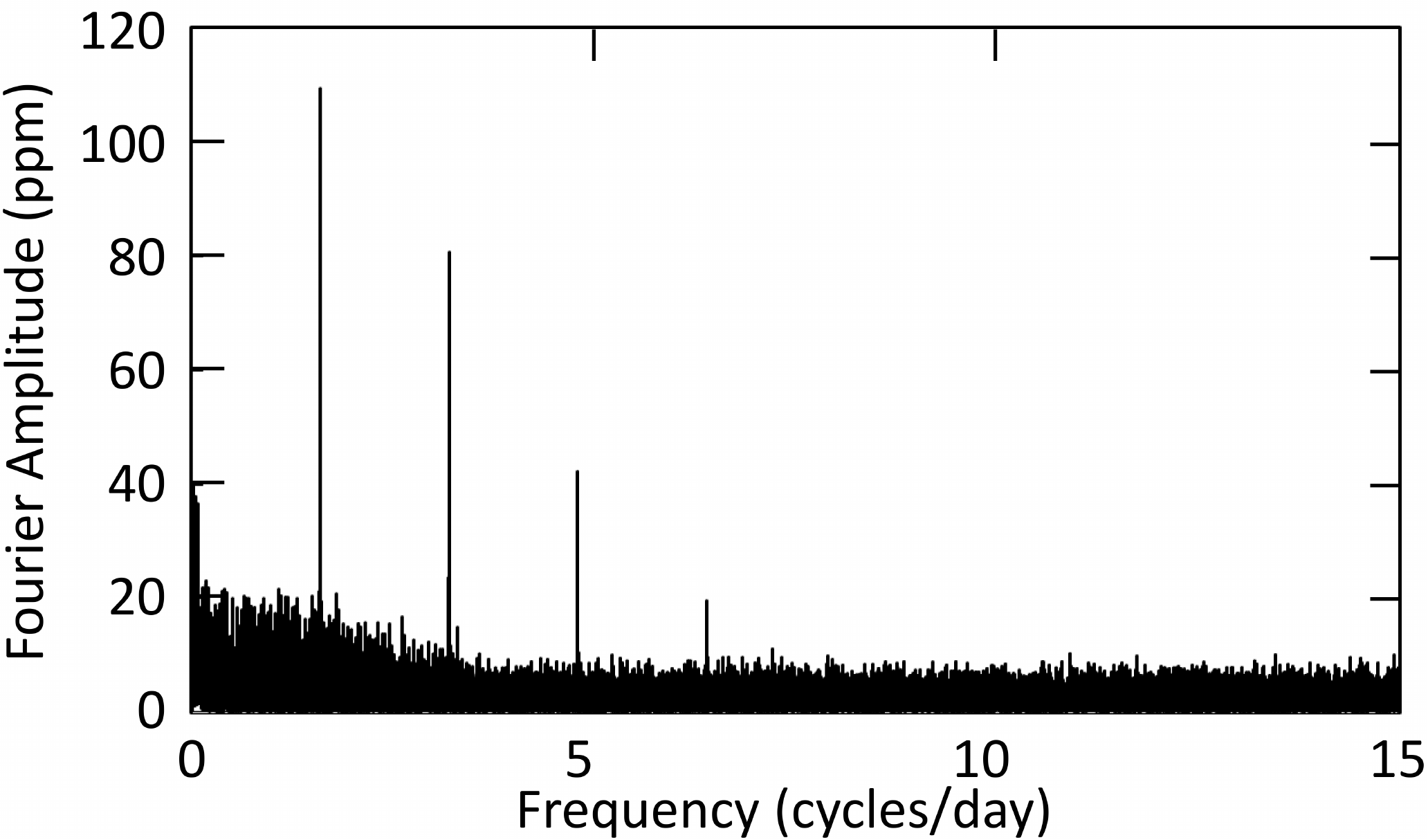}
\caption{Fourier transform of simulated forward scattering peaks superposed on an illustrative {\em Kepler} data set for a 15th magnitude star (see text for details). The peaks in the time domain were spaced every 2/3 days, and they had an amplitude of 300 ppm. The base frequency of 1.5 cycles/day and its following three harmonics are readily visible. The apparent excess (i.e., rising) background toward low frequencies is due to unresolved peaks of a 32-day planet in the system (see text).}
\label{fig:FT2}
\end{center}
\end{figure}

To demonstrate that such signals (with amplitudes of $\sim$100-500 ppm) are readily detectable in {\em Kepler}-quality data, we superposed a sequence of positive forward-scattering peaks of amplitude 300 ppm onto an actual {\em Kepler} photometric data set  (all 17 quarters) for a 15th magnitude star (KIC 11018648; KOI 759; Kepler 230). The results of the Fourier transform are shown in Figure \ref{fig:FT2}. This system also happens to have two planets in 32.6- and 91.8-day orbits. The signal from the 32.6-day planet ({\em Kepler} 230b) shows up with $\sim$100 closely packed harmonics in the region with frequencies $\lesssim 3$ cycles/day. {\em Kepler} 230c with $P_{\rm orb}=91.8$ days and a transit depth of 713 ppm is too low in frequency to appear in the FT in an obvious way. By contrast, the simulated added short-period planet with $P_{\rm orb} = 2/3$ days with its forward scattering peaks is readily detectable.  

The base frequency of 1.5 cycles/day is immediately evident, as are its next three higher harmonics. The FT amplitudes fall off roughly linearly with harmonic number for this particular example. Narrower forward-scattering peaks produce more higher harmonics, and vice versa.

Based on the three known dusty-tailed planets, and the fact that high equilibrium temperatures are expected at short orbital periods (more conducive to the release of dust), we expect that most of the dusty-tailed planets yet to be discovered will be found with $P_{\rm orb} \lesssim 1$ day. For such short-period systems, we have found that a simple Fourier analysis is just as effective a search tool as is the BLS algorithm (see Sanchis-Ojeda et al.~2014; and references therein), and is much faster.

\section{Searching for Forward Scattering Systems}
\label{sec:searches}

A comprehensive search through the main-field {\em Kepler} data set for ultrashort period planets (`USP's; exoplanets with orbital periods $\lesssim 1$ day) was carried out by Sanchis-Ojeda et al.~(2014) and yielded 106 such USP candidates. This is just the orbital period range that would be of interest for studying non-transiting exoplanets with dusty tails. The main difference in the proposed search would be to look for positive-going bumps in the light curve as opposed to negative-going transit dips. At the time of the Sanchis-Ojeda et al.~(2014) USP study, those carrying out the search were not aware of the possibility of finding narrow, positive-going bumps corresponding to non-transiting dusty-tailed planets. If the search had found the type of folded light curve illustrated in Fig.~\ref{fig:fs_peaks}, their significance would not have been recognized, and they would have been discarded with the other false positive targets. 
 
In the future, a comprehensive search for positive going bumps in short period systems, through the entire main {\em Kepler} field, the complete K2 data set (Howell et al.~2014), and the upcoming TESS data (Ricker et al.~2014) should be carried out. The criteria for a computer algorithm to search for forward scattering events would include: (1) A single positive-going symmetric peak in the folded light curve. (2) The full width at half maximum of this peak should be $\lesssim$20\% of $P_{\rm orb}$. (3) There should be no other significant peaks or minima in the folded light curve. (4) The amplitude of the peak should be $\lesssim 1000$ ppm in order to be plausibly ascribable to forward scattering. (5) An orbital period of $\lesssim 1$ day is expected. (6) For {\em Kepler} main field targets, the candidate dusty-tailed planet would also have to pass many of the same tests used to vet the regular planet candidates; e.g., light-centroid test, odd-even transit depth test, etc.~(see, e.g., Sanchis-Ojeda et al.~2014). 

There are two caveats worth mentioning about these selection criteria. First, if the dust tail is long, i.e., comparable to the size of the host star, then the dust scattering profile may violate the symmetry requirement. Second, the requirement that the peaks should be narrow, rather than sinusoidal in shape, depends on the assumption about the characteristic sizes of the scattering dust particles. In particular, if the particles are small, leading to broader scattering profiles, then the intensity of the forward scattering peak can be diminished to the point where it cannot be detected. Furthermore, such signals would tend to show up mainly at the base frequency in Fourier transforms. Since there are many stellar phenomena whose signature matches a low-amplitude sinusoid, it would be difficult to distinguish a dust emitting planet from various false alarms.

We are currently engaged in systematically searching through the entire K1 data set for such non-transiting exoplanets with dust tails.  However, this search will not be completed for some time now and, in any case, is beyond the scope of the present work.  We will report on the results of our search elsewhere.

\section{Summary and Conclusions}
\label{sec:concl}

In this work we have proposed a novel approach to detecting non-transiting dust emitting planets via their positive-going forward-scattering peaks. We have calculated in detail, partially analytically, what these scattering peaks should look like. Scaled to the few transiting planets we know of with dusty tails, it appears quite realistic that the amplitudes of the scattering peaks could be in the range 50-500 ppm. Furthermore, because these can be detected to orbital inclination angles of $\gtrsim 45^\circ$, we might potentially be able to detect about an equal number of non-transiting as transiting dusty planets via this method. 

The main message for researchers searching for planets is to look carefully at objects that have only positive-going signals with periods of $\lesssim$ 1 day before discarding them as false positives. Both K2 and the upcoming TESS mission should be quite sensitive to the detection of such short period planets.

Detection of non-transiting dusty-tailed planets would provide important clues about the dust size. If the particles are too large, then the required scattering angles of $\gtrsim 90^\circ - i$ will not be attained. Conversely, if the dust particles are too small, then the scattered radiation will be dispersed over too wide a range of angles (e.g., $\gtrsim 60^\circ$), and the forward scattering peaks will be both reduced in amplitude and their FT's will not contain sufficient harmonics to allow them to be distinguished from other astrophysical phenomena. Putting these requirements together with the simple expression for characteristic scattering angles in Eqn.~(\ref{eqn:angle}), we find a basic requirement for the grain size of:
\begin{equation}
\frac{1}{3} \mu{\rm m} ~\lesssim ~a ~\lesssim \frac{20^\circ}{90^\circ -i} ~ \sim \frac{2}{3} {\mu{\rm m}}
\end{equation}
For a distribution of grain sizes, the distribution would have to include a substantial fraction of grains in this range. 

Thus far, detected dusty-tailed planets are rare. Therefore, even the prospects for roughly doubling this sample seems quite attractive.

\section*{acknowledgement}

This work was supported in part by NASA/HST grant GO 12987.01-A.








\appendix

\section{Calculation of the Effective Phase Function}
\label{app:ptheta}

\subsection{Geometry}
\label{app:def}

The geometry of the scattering problem is defined in Figure \ref{fig:geometry}. The $\hat{z}$ axis is defined by the line that connects the center of the host star to the scattering dust cloud, the latter of which is assumed to be small in size compared to the star. A photon leaving the star and heading for the scattering cloud makes an angle $\gamma$ with respect to the {$\hat{z}$ axis. The scattered photon is assumed to lie in the $\hat{x}-\hat{z}$ plane without loss of generality because of the cylindrical symmetry of the problem. The photon heading toward the scattering cloud is {\em not} assumed to lie in that plane. The goal is to find the `effective phase function', $\mathcal{P}(\Theta)$, of the integrated outgoing scattered radiation pattern as a function of the angle $\Theta$, which is symmetric with respect to the $\hat{z}$ direction.

The specific intensity at the stellar surface, $I_\nu(\Phi)$, is the power radiated per frequency interval at $\nu$ per square centimeter of surface area, per unit solid angle at angle $\Phi$ with respect to the normal (defined by $\hat{n}$). We drop the subscript $\nu$ for simplicity. The power emitted through a surface area of the star, $\Delta A_s$, going into a solid angle $\Delta \Omega$ at an angle to the normal, $\Phi$, is $\Delta P =  I(\Phi) \, \Delta A_s \, \Delta \Omega \, \cos \Phi$. 

Consider a small spherical scattering cloud at distance $d$ with projected area $A_c$ as seen from far away. The radius of the star is $R_s$. The observer is at a far distance $D$, and is using a detector of area $A_d$. The star has luminosity $L_s$ which is given by $I(0) \mathcal{I} \pi (4 \pi R_s^2)$ where $\mathcal{I} \equiv 2 \int_0^{\pi/2} [I(\Phi)/I(0)] \sin \Phi \cos \Phi d\Phi$, and where in turn, $I(\Phi)/I(0)$ is the limb-darkening function for the star. $\mathcal{I}$ is defined so that for a Lambertian emitter this constant is 1. We can then write the full flux of the star at the Earth as $F_0=L_s/(4 \pi D^2) = \pi  I(0) \mathcal{I} R_s^2/ D^2$. 

\subsection{Scattering Out of the Beam}
\label{app:scatt_out}

The power deficit due to a transit in a detector at the Earth from an area $\Delta A_s$ near mid-transit is
\begin{equation}
\delta P_{\rm so} ~=~ I(0) \, \mathcal{I} \,\Delta A_s \, \frac{A_d}{D^2}\, \left(1-e^{-\tau}\right) ~=~ I(0)\, \mathcal{I} \, A_c \, \frac{A_d}{D^2}\, \left(1-e^{-\tau}\right)
\end{equation}
where the last term is the fraction of the photons that leave the beam by scattering at least once, and $\tau$ is an appropriate area-averaged optical extinction thickness of the cloud. The ratio $A_d/D^2$ is the solid angle of an Earth-based detector as seen by an element on the surface of the star, and the relevant area of the star which will contribute to the transit is $\Delta A_s \equiv A_c$ (the projected area of the cloud). 
To determine the fraction of the flux at the Earth which has been removed due to the transit, we divide by the area of the detector, $A_d$, and normalize this to the total flux of the star.
\begin{equation}
\frac{\delta F_{\rm so}}{F_0} ~=~ \frac{I(0) \,\mathcal{I}}{F_0} \, \frac{A_c}{D^2}\, \left(1-e^{-\tau}\right) ~=~ \frac{A_c}{\pi R_s^2}\, \left(1-e^{-\tau}\right)
\label{eqn:scatt_out}
\end{equation}

\begin{figure*}
\begin{center}
\includegraphics[width=0.7 \textwidth]{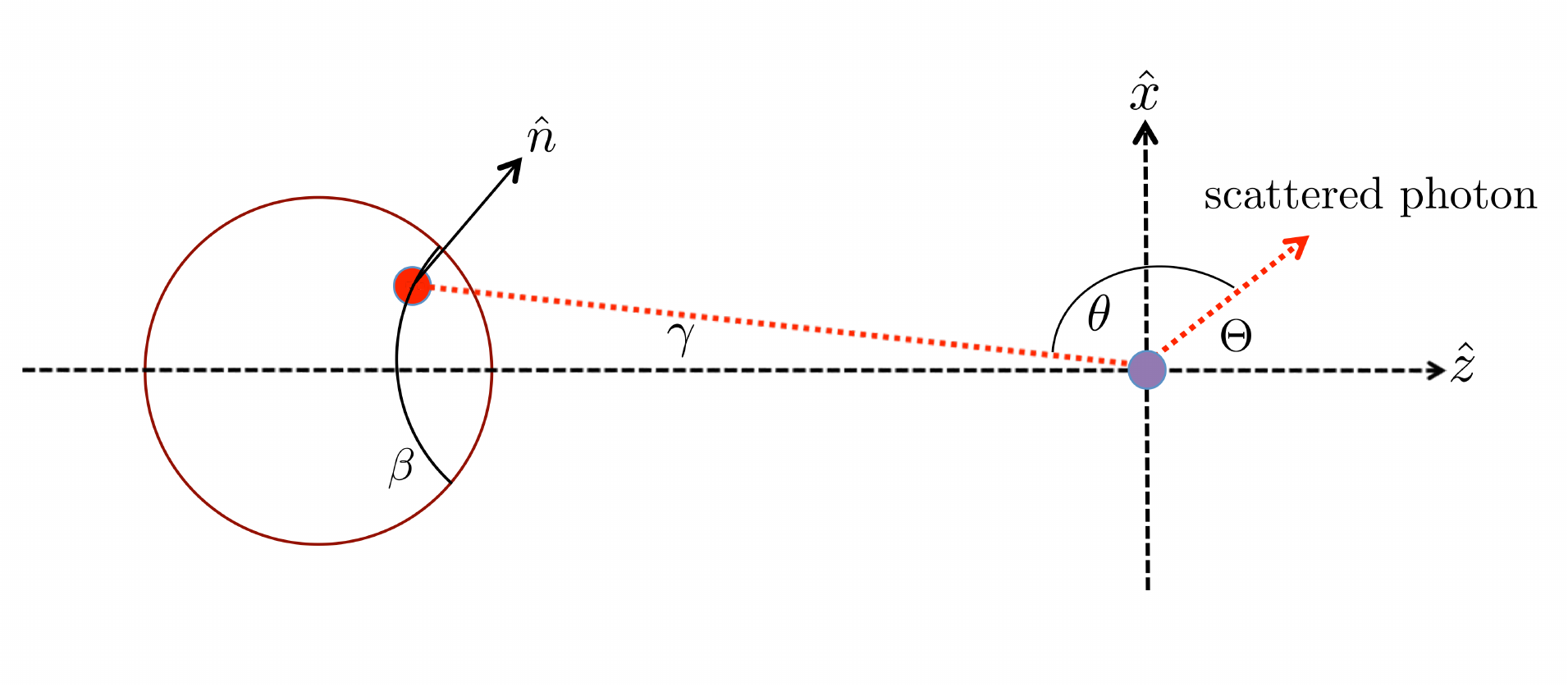}
\caption{Scattering geometry used in this work. The star is shown on the left and a small spherical dust scattering cloud is on the right (purple). A photon (dotted red line) leaves a small emitting region (red patch) on the star, travels to the scattering cloud, and scatters through an angle $\theta$. Because of the cylindrical symmetry of the problem, we rotate the $\hat{x}$ axis around the $\hat{z}$ axis until the scattered photon lies in the $\hat{x}-\hat{z}$ plane. The angle that the scattered photon makes with the $\hat{z}$ axis is $\Theta$. The angle between the direction of the photon from the star to the scattering sphere and the $\hat{z}$ direction is $\gamma$. $\hat{n}$ marks the normal to the surface area of the emitting patch. Finally, $\beta$ is the angle around a small circle whose pole is the $\hat{z}$ direction.}
\label{fig:geometry}
\end{center}
\end{figure*}

\subsection{Scattering Into the Beam}
\label{app:scatt_in}

Consider a small surface area on the star that is radiating into the dust scattering cloud. This radiation is characterized by an angle $\Phi$ with respect to the local normal, $\hat{n}$. The power into the cloud from that patch of surface area is:
\begin{equation}
\delta P_{\rm si} ~=~ I(\Phi) \,\Delta A_s \cos \Phi \,d\Omega_c ~=~I(\Phi) \,\Delta A_s \cos \Phi \,\frac{A_c}{r^2}
\end{equation}
where $d\Omega_c$ is the solid angle of the scattering cloud as seen from that patch of the star's surface, and $r$ is the distance from the stellar patch to the scattering cloud. 
By the reciprocal nature of the specific intensity, we can also compute the power received by the cloud using the solid angle as viewed from the scattering cloud, $d\Omega'$, and the projected area of the scattering cloud, as seen by a photon coming in from the star: $I(\Phi) \,A_c \,d\Omega'$. Conveniently, the scattering cloud is taken to be a sphere, so the projected cloud area is always $A_c$, regardless of the incident angle. Therefore, the power scattered in the dust cloud and then into a detector on the Earth can be written as:
\begin{equation}
\delta P_{\rm si} = I(\Phi) \,A_c \,d\Omega' \, \varpi  \tau e^{-\tau}\,P(\theta) \,\frac{A_d}{D^2}
\label{eqn:dscatt_in}
\end{equation}
where $A_d$ and $D$ have the same meaning as above, and $\varpi$ is the single scattering albedo (the ratio of the scattering to the extinction cross section; see e.g., Liou 2002). For the present calculations we set $\varpi = 1$. $P(\theta)$ is defined as the probability per unit solid angle of dust scattering by an angle $\theta$, and is normalized to unity. The factor $\varpi \, \tau \, {\rm exp}^{-\tau}$ is the probability of the radiation scattering {\em just once} in the dust cloud.

In order to integrate the contribution from the entire facing hemisphere of the star we define a number of different angles in Fig.~\ref{fig:geometry}:
$\gamma$ is the angle between a photon moving toward the scattering cloud and $\hat{z}$;
$\beta$ is the azimuthal angle around the line defined by $\hat{z}$; 
$\theta$ is the scattering angle for light leaving the star, scattering, and then going toward the Earth;
$\Theta$ is the angle between the final scattered direction and $\hat{z}$; and
$\Phi$ is the angle between the local normal to the stellar surface, $\hat{n}$, and the direction to the scattering cloud.
The $\hat{x}$ direction is chosen such that the final scattered photon travels in the $\hat{x}-\hat{z}$ plane. Since the final scattering around the line joining the star and dust cloud must be axisymmetric, there is no loss in generality by choosing the final scattering plane in this way. Finally, $\hat{y}$ is the remaining orthogonal direction. A point on the star with $\beta = 0$ lies in the $\hat{x}-\hat{z}$ plane, while a point at $\beta = 90^\circ$ lies in the $\hat{y}-\hat{z}$ plane.

We now define two important unit vectors. 
The vector describing the final direction of the scattered photon is:
\begin{equation}
\hat{V}_s = \sin \Theta \,\hat{x}+\cos \Theta \, \hat{z}
\end{equation}
The unit vector of a photon traveling from the stellar surface to the scattering cloud is:
\begin{equation}
\hat{V_p}=-\sin \gamma \cos \beta \, \hat{x} - \sin \gamma \sin \beta \,\hat{y} +\cos \gamma \, \hat{z}
\end{equation}
Now the actual scattering angle $\theta$ can be defined in terms of $\Theta$ (i.e., with respect to the line joining the star and cloud) in terms of:
\begin{equation}
\cos \theta = \hat{V}_p \cdot \hat{V}_s = - \sin \gamma \cos \beta \sin \Theta +\cos \gamma\,\cos \Theta
\end{equation}

We can now integrate Eqn.~(\ref{eqn:dscatt_in}) over the facing hemisphere of the star, by taking the solid angle, $d\Omega' = \sin \gamma \, d\gamma \,d\beta$.
This leads to the total power scattered into the distant detector:
\begin{equation}
\Delta P_{\rm si} =  \frac{A_d A_c}{D^2} \, \varpi \tau e^{-\tau} \int_0^{2 \pi} \int_0^{\gamma_s} I(\Phi) \,  \sin \gamma \,P(\theta) \, \, d\gamma \, d\beta
\end{equation}
where $\gamma_s =\sin^{-1} (R_s/d)$.

The flux scattered into the detector is given by the power $\Delta P_{\rm si}$ divided by $A_d$.  If we form this ratio and then divide by $F_0$ to find:
\begin{equation}
\frac{\delta F_{\rm si}}{F_0} \simeq \,\frac{A_c}{\pi R_s^2}\, \varpi \tau e^{-\tau} \int_0^{2 \pi} \int_0^{\gamma_*} I'(\Phi) \, \sin \gamma  \,\,P(\theta) \, d\gamma \, d\beta
\label{eqn:scatt_in}
\end{equation}
where $I'(\Phi)$ is the specific intensity of the star normalized to the quantity $\mathcal{I} \,I(0)$.

\subsection{Ratio of Scattering Into the Beam vs.~Scattering Out of the Beam}
\label{app:ratio}

Let us now divide equation (\ref{eqn:scatt_in}) by equation (\ref{eqn:scatt_out}) to find the ratio of the scattering amplitude compared to the transit depth (as would be seen by a distant observer in the orbital plane):
\begin{equation}
\begin{split}
& \frac{\rm Scattering~Amplitude}{\rm Transit~Depth} \simeq \\
& \frac{\varpi \tau e^{-\tau}}{\left(1-e^{-\tau}\right)} \int_0^{2\pi}\int_0^{\gamma_s} I'(\Phi) \, \sin \gamma   \,\,P[\theta(\gamma,\beta; \Theta)] \, d\gamma \, d\beta
\end{split}
\end{equation}
where, to recap, the relations between $\gamma$ and $\Phi$, and between $\theta$ and $\Theta$, are:
\begin{equation*}
\sin \Phi = \frac{d}{R_s}\, \sin \gamma  \, ~ = \frac{ \sin \gamma }{ \sin \gamma_s }  
\end{equation*}
\begin{equation*}
\cos \theta = - \sin \gamma \cos \beta \sin \Theta +\cos \gamma\,\cos \Theta
\end{equation*}

We can write $I'(\Phi)$ in terms of a generalized polynomial limb-darkening function for the star:
\begin{equation}
I'(\Phi) = 1 -\sum \,\alpha_n \left(1-\sqrt{1-\sin^2 \gamma/\sin^2 \gamma_s}\right)^n
\label{eqn:LD}
\end{equation}

Finally, we write the ratio of scattering into vs.~scattering out of the beam as:
\begin{equation}
\frac{\rm Scattering~Amplitude}{\rm Transit~Depth} \simeq \pi \left(\frac{R_s}{d}\right)^2\frac{\varpi \tau e^{-\tau}}{\left(1-e^{-\tau}\right)} \, \mathcal{P}(\Theta)
\label{eqn:ratio}
\end{equation}
where the effective phase function $\mathcal{P}(\Theta)$ is given by:
\begin{equation}
\mathcal{P}(\Theta) = \int_0^{2\pi}\int_0^{\gamma_s} \frac{I'(\Phi) \, \sin \gamma   \,\,P[\theta(\gamma,\beta; \Theta)] \, d\gamma \, d\beta}{\pi \sin^2\gamma_s}
\label{eqn:EPF}
\end{equation}
In the limit of a small star and/or a large cloud-star separation, $\mathcal{P}(\Theta) \rightarrow P(\theta)$.  

The dependence of the ratio of scattering into vs.~out of the beam on the separation of the cloud and star now appears explicitly by the leading factor $(R_s/d)^2$, and is not involved in the effective phase function (see Eqn.~\ref{eqn:EPF}).

\begin{figure*}
\begin{center}
\includegraphics[width=0.85 \textwidth]{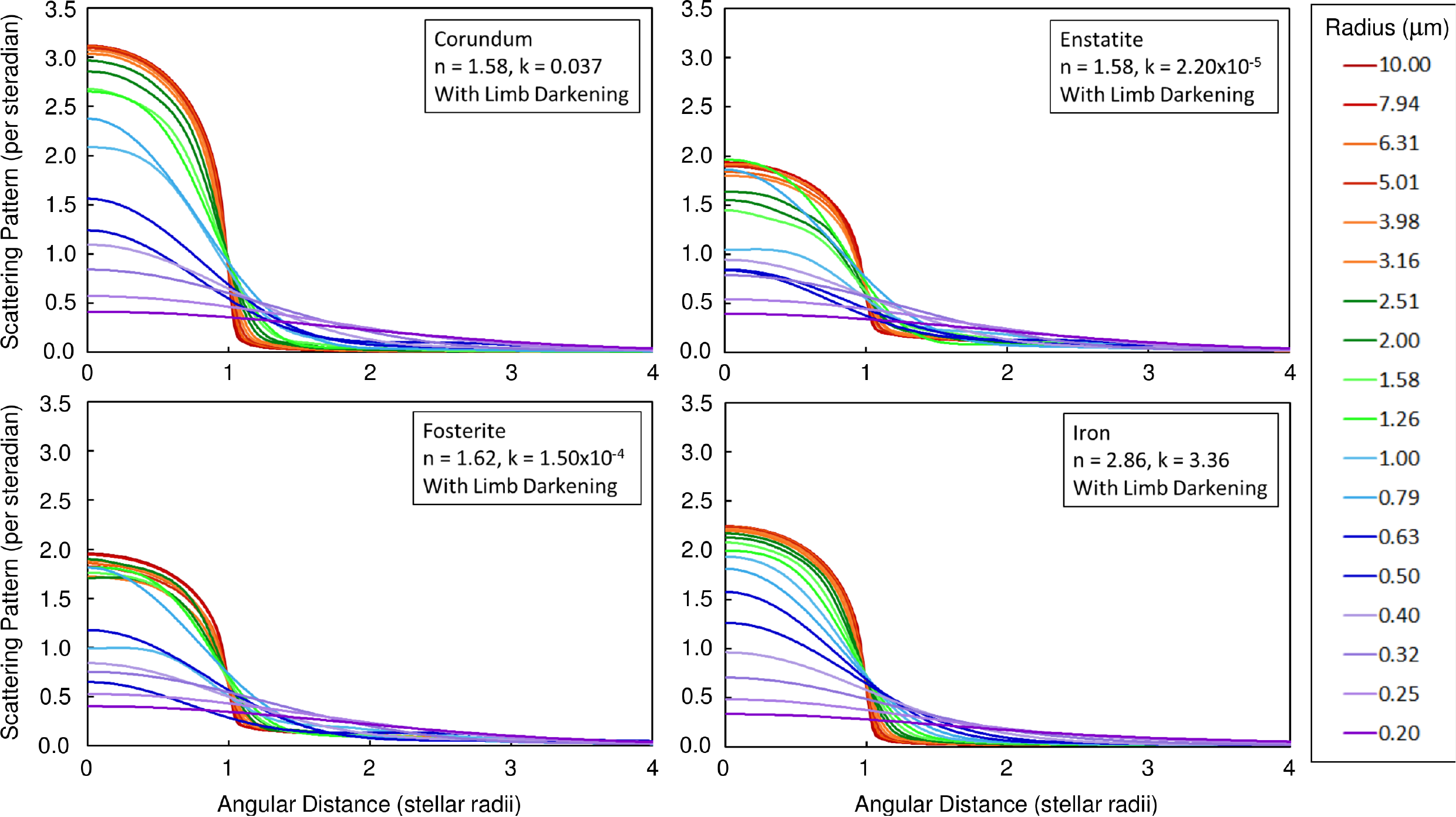}
\caption{ Effective phase functions calculated according to Eqn. \ref{eqn:EPF} based on particle phase functions calculated using Mie scattering theory. The particle radii are shown in the legend, while the indices of refraction for the four minerals are shown in the panel insets, where n and k are the real component imaginary components. Note the modest variability in the peak heights and the angular distributions of the effective phase functions for a given radius depending primarily upon the imaginary component of the index of refraction k. }
\label{fig:EPFmontage}
\end{center}
\end{figure*}

\subsection{Results for Effective Phase Functions}
\label{app:EPF}

Equation (\ref{eqn:EPF}) is the integral that we used to generate the effective phase functions in this paper.  We show four illustrative sets of such curves in Fig.~\ref{fig:EPFmontage} for the minerals corundum, enstatite, forsterite, and iron.  The curves are for 18 different particle sizes ranging from 0.2 to 10 $\mu$m in logarithmic steps. The ``observing'' wavelength is fixed at 0.55 $\mu$m.  The illustrative quadratic limb darkening coefficients used in Eqn.~(\ref{eqn:LD}) are $a_1 = 0.467$ and $a_2 = 0.277$. All of the particle phase functions, $P(\theta)$, were computed from the Bohren \& Huffman (1993) Mie scattering code.

The differences among the various sets of curves for a given radius result largely from the values of $k$, the imaginary part of the index of refraction (leading to absorption), among the different minerals. This leads to two effects that are mineral specific.  In the first, the peak values of the effective phase functions (i.e., at $\Theta = 0$) vary by $\sim \pm 25\%$.  Second, the ordering of the curve amplitudes with particle size is not strictly monotonic due to the details of the Mie scattering for particles with $a \sim 2\pi \lambda$.  The overall trend, however, is quite clear where smaller particles produce broad scattering patterns, while large particles yield an effective phase function that traces out the angular profile of the host star.

To illustrate the general trend of these curves with particle size, without the irregular progressions seen for some of the minerals, we used the same Eqn.~(\ref{eqn:EPF}) to generate Fig.~\ref{fig:phasefunc} in the main body of the text, but there we substituted simple Airy functions for the more complex Mie scattering patterns. In the absence of knowing the detailed grain material in the putative dust tails, or their size distribution, it is easier to consider these illustrative effective phase functions without the complications that are specific to different minerals.


\bsp	
\label{lastpage}
\end{document}